**Analyzing Blood Glucose Levels with Near Infra-Red Spectroscopy and Chemometric Multivariate Methods**

Hadi Barati[1], Arian Mousavi Madani[1], Soheil Moradi[1], and Mehdi Fardmanesh[1]

[1]Department of Electrical Engineering, Sharif University of Technology, Tehran, Iran

**Corresponding author:**
Hadi Barati
Department of Electrical Engineering, Sharif University of Technology, Tehran, Iran
Phone: +989369923044
Email: baratihadi6@gmail.com , hadi.barati@ee.sharif.edu

**Abstract**
In this work, the blood NIR absorbances are recorded using the FT-IR method. It is shown that when the absorbance curves are multiplied by the first derivative of the water absorbance spectrum as well as by the first derivative of the glucose absorbance, the peaks related to the water interferent in the blood are effectively removed from the blood absorbance spectra, allowing for better distinction of the peaks of the blood glucose. The PCR prediction using this method shows smaller errors compared to the PCR employing the net absorbances, while the number of derived principal components is smaller in the PCR method based on the derivatives than the one based on the net absorbances. Additionally, the prediction of blood glucose levels using a linear regression model based on the molar absorptivity of glucose also demonstrates acceptable accuracy.

**Introduction**

Diabetes is a metabolic disorder characterized by persistent hyperglycemia and its complications, which generally place a significant burden on healthcare systems. Conventional methods of blood glucose measurement are invasive finger-prick tests that usually cause discomfort, resulting in low compliance and are not easily accessible, especially where resources are limited. These disadvantages have stimulated widespread research efforts toward developing non-invasive glucose monitoring technologies designed to help patients follow their treatment plans.[1-8] Of these, near-infrared spectroscopy is one of the most promising approaches in non-invasive glucose monitoring because it has the unique ability to penetrate biological tissues and capture molecular vibrations. Compared to MIR and Raman spectroscopy, Near Infra-Red (NIR) spectroscopy[9] requires minimal sample preparation. It also has a greater depth of penetration and is suitable for bulk analysis, making it very attractive for biomedical applications.[10] However, the drawbacks of this approach include overlay spectral bands of sample components as well as great sensitivity to environmental variables such as temperature when trying to extend its use. In addition, demanding data processing methods are required for chemometric analysis of the measured data. Recent advances in computational modeling have overcome some of these challenges. Techniques such as multilevel principal component regression (MLPCR) effectively correct for temperature-induced spectral changes, making NIR-based predictions more robust under changing environmental conditions. Machine learning methods, including Support Vector Machine Regression (SVMR) and hybrid Principal Component Analysis-Neural Networks (PCA-NN), have also achieved very good predictive performance by identifying key spectral features among complex and noisy data.[11] Furthermore, chemometric models have significantly improved the reliability of glucose measurements through new methodologies. Optimized multivariate calibration methods[12], such as Partial Least Squares Regression[13], can handle nonlinearities in spectral data to enhance prediction accuracy in glucose measurement.[14-16] Complementary approaches, such as classification-before-regression frameworks, have further improved glucose predictions by segmenting data into glycemic categories before the application of regression models.[17] While design innovation in sensors and signal processing has improved the practicality of the measurement[18], the use of dual-spectral photoplethysmography signals, when combined with the analysis of heart rate variability, has resulted in highly accurate and clinically acceptable glucose monitoring over a large range of glucose concentrations.[19] Fingernail glycated keratin is one of several alternative substrates that have been tried with alternative approaches in pursuit of non-invasive screening and monitoring.[20] Despite these successes, a number of challenges remain, especially in applying in vitro successes to more reliable in vivo applications. Environmental and physiological interference, such as temperature variation and individual variability, are still challenges.[21] These will require innovative hardware solutions and sophisticated computational models to achieve results that are reliable, reproducible, and clinically applicable.

In this work, the blood glucose levels of twenty healthy subjects will be measured using a commercial glucometer and also be estimated using chemometric multivariate methods on the measured infrared absorption of the collected blood samples. New approaches are proposed to effectively cancel out the water interfering absorption and isolating the absorption peaks that are attributed to glucose. The present paper is organized as follows: in the Mathematical section, the computational methods for estimating the blood glucose levels from the measured transmittance are described. In the Experimental section, the empirical method for collecting the blood samples and measuring their glucose levels are explained. The obtained results are discussed in the Results and Discussion section. Concluding remarks are provided in the Conclusion section.

**Mathematical method**

Principal Component Regression

Principal component regression (PCR) is a chemometric multivariate analysis method in which the concepts of Principal Component Analysis (PCA) and multiple linear regression are combined to extract linear relationships between the predictor variables and the responses. Here, the predictor variables are the blood infrared absorbances measured (calculated) at various wavenumbers, and their corresponding principal components (PCs) obtained by PCA. On the other hand, the responses are the blood glucose concentrations. The stages of PCR include: 1. Data filtering and standardization to ensure that the variables being analyzed are noiseless and on a similar scale. To standardize the data, the mean value is subtracted from each data point and the result is then divided by the data standard deviation. 2. Applying

PCA to transform variables into orthogonal components capturing maximum variance. 3. Developing a regression model between PCs and responses while determining the optimum number of PCs through cross-validation techniques. 4. Assessment of the PCR model employing metrics like Root Mean Squared Error (RMSE) to investigate the effectiveness of the model's predictive power.

In this work, the measured absorbances are stored as a data matrix denoted by $\mathbf{A}_{n \times m}$. n and m are the number of predictor variables and the number of blood samples, respectively. After applying PCA, the dimensionally reduced data matrix can be presented as $\mathbf{A}'_{n \times m'}$. $m'$ is the number of PCs. $\mathbf{A}'$ contains the projected absorbances onto PCs. Thus, applying PCR on the data yields a linear relationship between PCs and the blood glucose levels as $\mathbf{c}'_{n \times 1} = \mathbf{A}'_{n \times m'} \times \delta'_{m' \times 1}$. $\delta'_{m' \times 1}$ is the regression coefficients. $\mathbf{c}'_{n \times 1}$ is a vector containing the estimated glucose levels. The comparison between these estimated (predicted) concentrations and the actual (measured) blood glucose levels denoted by $\mathbf{c}_{n \times 1}$ exhibits the accuracy of the PCR method.

Absorptivity Regression Analysis

As a complementary method for predicting glucose levels, a novel regression method[22] is employed in the present work. This method, called Absorptivity Regression Analysis (ARA), has been developed by an ideal combination of linear regression concept and Lambert-Beer's relationship. The Lambert-Beer relationship, which has been empirically validated, shows a linear relationship between the measured absorbance and the concentration of the solute in an irradiated solution. This equation is presented as $\mathbf{A}_{n \times m} = \mathbf{c}_{n \times 1} \eta_{1 \times m} l$. $\eta_{1 \times m}$ is the molar absorptivity of the solute like glucose. $l$ is the thickness of the sample solution. As previously explained in details,[22] the Lambert-Beer equation can be simplified to an absorptivity-concentration relationship presented as $\eta_{1 \times m} \times \delta_{m \times 1} = l^{-1}$. $\delta_{m \times 1}$ is the regression coefficients of this linear equation and can be determined through equation $\sum_{i=1}^{m} \eta_i \delta_i = \tilde{\upsilon}_{PL}$. $\tilde{\upsilon}_{PL} \equiv l^{-1}$ which is defined as the wavenumber corresponding to the pathlength. $\eta_i$ is the value of molar absorptivity at wavenumber $\tilde{\upsilon}_i$. As suggested in our previous work[22], $\delta_i$ is arbitrarily defined by the following equations

$$\delta_i = \frac{\tilde{\upsilon}_i}{\eta_i} \quad i = 1, 2, \ldots, m-1$$

$$\delta_m = \frac{1}{\varepsilon_m} \left( \tilde{\upsilon}_{PL} - \sum_{i=1}^{m-1} \eta_i \delta_i \right)$$

(1)

This is a blind method for calculating the regression coefficient $\delta_i$ using equations (1). Hence, it is natural to have errors in the estimation of the glucose concentrations $\mathbf{c}_{n \times 1}$. The error originates from the uncertainty in measurements of the absorbances and the molar absorptivity. In other words, the equation $\sum_{i=1}^{m} \eta_i \delta_i = \tilde{\upsilon}_{PL}$ presents an infinite number of solutions for $\delta_i$. The solutions delivered by Eq. (1), is one of these solutions and may not fit the measurement. In our previous work[22], by chance, the ARA solutions matched the experimental results with acceptable accuracy. However, In the current study, it has been noted that the ARA in its original form, do not provide accurate estimation of glucose concentrations. As a result, we suggest incorporating a term denoted by $\mathbf{R}_{n \times 1}$, for error compensation for the regression equation, namely $\mathbf{c}_{n \times 1} = \mathbf{A}_{n \times m} \times \delta_{m \times 1} + \mathbf{R}_{n \times 1}$. $\mathbf{R}_{n \times 1}$ is the error vector and can be defined in such a way that the errors introduced to the predictive model due to errors in the FTIR measurements, become the least. We will see that by appropriate adjustment of this factor, prediction by ARA can become as accurate as possible. In the current study, the Python package is utilized for preprocessing the measured data as well as the PCR analysis.

**Experimental method**

Twenty healthy male and female subjects participated in the experiment of the present study. After fasting overnight, their blood samples were collected after receiving breakfast and lunch. The glucose level of each sample was assessed by a commercial finger-prick glucometer. On the other hand, the Bruker's Vertex 70 FT-IR spectrometer was employed to measure the transmittances of blood samples. The measurement wavenumber range was set at 4000–8000 cm$^{-1}$ (NIR range anticipated for the maximum IR absorption by glucose[22]) with a scanning step of 1 cm$^{-1}$. Compared to other methods like finger-pricking and subcutaneous sensor methods, the FT-IR approach can be a more reliable and cost-effective choice for glucose monitoring due to its noninvasive nature and precise results. The measurements were repeated on five consecutive days, with each subject providing two samples each day, leading to a total of 200 blood samples collected, with the first measurement taken 30 minutes post-breakfast and the second one taken 30 minutes post-lunch. To ensure accurate measurements, the liquid cell used for collecting the blood sample was thoroughly cleaned with ethanol and deionized water before each sample preparation. Finger-pricked blood drops were immediately encapsulated in the liquid cell to prevent contamination risks from air and skin. The use of a concentrated and pure sample in a liquid cell improves the sensitivity and specificity of FT-IR measurements.

**Results and discussion**

Absorbances have calculated from the measured transmittances using equation $\mathbf{A} = -\log \mathbf{T}$. $\mathbf{T}$ is the measured sample transmittance. In Figure 1, some of the measured transmittances of the 200 blood samples are illustrated. The glucose concentration level, in [mg/dL], of these samples are also shown. The transmittance of the pure water measured by the same FTIR device is drawn, too.

For the chemometric analysis, the net absorbances are obtained by subtracting the pure water absorbance from the sample absorbance. The water absorbance is determined from the measured transmittance of the water. Although several blood components contribute to the sample absorbance, in the present work, only the water contribution is taken into account because other interferents, such as protein and lipid molecules, have much less effect on the infrared transmittance spectrum compared to water with very high IR absorbance.[22-27] In other words, after comparing the blood transmittance curves with that of water in Figure 1, it is clear that water is a significant interferent substance in FTIR blood glucose spectroscopy. Thus, its contribution must be removed from the data as the first step prior to any further data processing in the purpose of accurate glucose predictions.

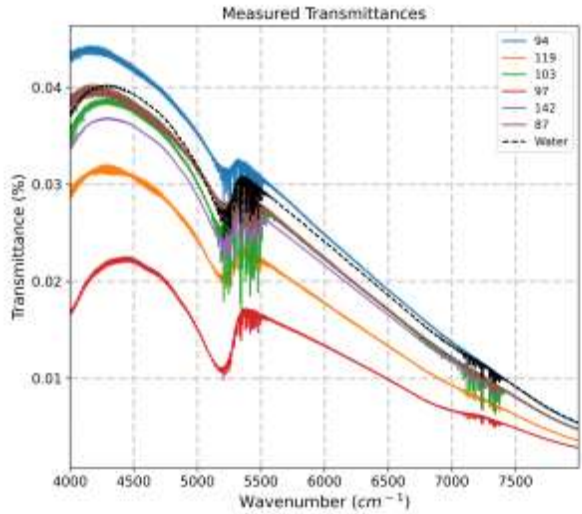

Figure 1. Measured transmittances. The glucose level of each sample has been shown in [mg/dL]. The water transmittance has also been drawn.

As previously devised,[28] a scale of the water absorbance is subtracted from the original absorbances, i.e. $\mathbf{A}_{net} = \mathbf{A} - \gamma \mathbf{A}_{water}$ in such a way that the water effect is canceled as much as possible. However, determining this scaling factor, i.e. $\gamma$, is a challenge because the amount of water in the blood sample varies from sample to sample, and thus, applying a unique scaling factor will not be effective in water cancellation. The effect of water removal is illustrated in Figure 2 for various values of $\gamma$. The obtained net absorbances were numerically filtered to remove the noises by employing Butterworth lowpass filter in Python software. As can be seen, the obtained net absorbances are different for different values of the scaling factor and therefore, the results being generated by the regression model will be sensitive to this factor. As $\gamma$ increases, the number of absorbance curves moving toward negative values increases. This means that the amount of IR absorbed by water in some samples was larger than that of the pure water sample. For $\gamma = 0.5, 0.83$, all absorbances take positive values across the entire wavenumber range while for $\gamma = 1$, some of the net absorbances take negative values at the entire wavenumber range. Since the exact amount of water in each blood sample was unknown, this approach clearly introduces errors into the prediction.

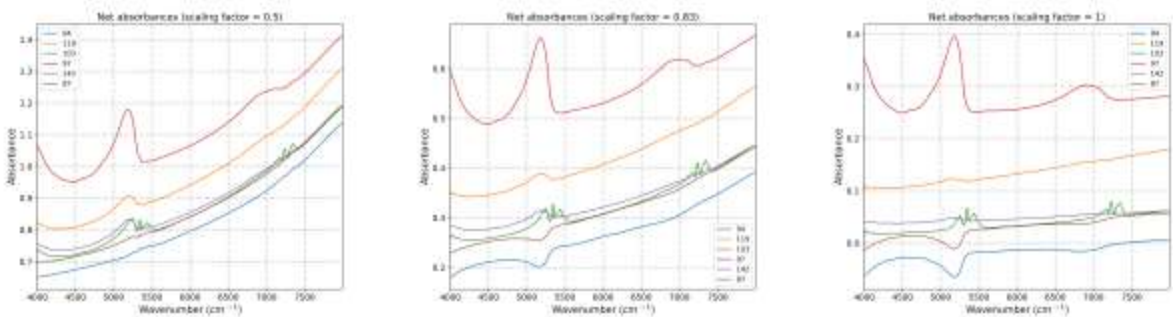

Figure 2. The effect of various values of the scaling factor on the net absorbances ($\mathbf{A}_{net}$). The glucose level of each sample has been shown in [mg/dL].

In this work, to alleviate this problem, three methods are proposed that the third one is obtained by the combination of the two other methods. In the first technique, it is suggested that each original sample absorbance be multiplied by the first derivative of the water absorbance with respect to wavenumber $\tilde{\upsilon}$. By using this approach, it is expected to effectively suppress the absorbance peaks of water due to derivative vanishment at the peaks, leading to more distinguishable glucose peaks in the obtained absorbance spectrum. This method which is called Derivative Multiplication Suppression (DMS) technique can be formulated as $\mathbf{A}_{DMS} = \mathbf{A} \cdot \frac{d\mathbf{A}_{water}}{d\tilde{\upsilon}} \cdot \left\| \frac{d\mathbf{A}_{water}}{d\tilde{\upsilon}} \right\|^{-1}$. The normalization term $\left\| \frac{d\mathbf{A}_{water}}{d\tilde{\upsilon}} \right\|^{-1}$ is necessary to soften the values of $\frac{d\mathbf{A}_{water}}{d\tilde{\upsilon}}$ and avoid harsh variations imposing on $\mathbf{A}_{DMS}$. $\| \ \|$ is the Euclidean norm also known as the L2 norm operator.

In the second method, the first derivative of glucose absorbance is calculated, and the term (1-glucose absorbance derivative) is multiplied by the original absorbance. This way, the sample absorbance peaks related to the glucose IR absorption will be strengthened. This method is called Derivative Multiplication Reinforcement (DMR) and is formulated as $\mathbf{A}_{DMR} = \mathbf{A} \cdot (1 - \frac{d\mathbf{A}_{glucose}}{d\tilde{\upsilon}}) \cdot \left\| 1 - \frac{d\mathbf{A}_{glucose}}{d\tilde{\upsilon}} \right\|^{-1}$.

In the third technique, DMS and DMR are simultaneously applied, i.e., $\mathbf{A}_{DMSR} = \mathbf{A} \cdot \frac{d\mathbf{A}_{water}}{d\tilde{\upsilon}} \cdot \left\| \frac{d\mathbf{A}_{water}}{d\tilde{\upsilon}} \right\|^{-1} \cdot (1 - \frac{d\mathbf{A}_{glucose}}{d\tilde{\upsilon}}) \cdot \left\| 1 - \frac{d\mathbf{A}_{glucose}}{d\tilde{\upsilon}} \right\|^{-1}$.

In figure 3, the pure water absorbance, its first derivative and the resulted $\mathbf{A}_{DMS}$ have been depicted. As can be seen, the peak due to water at 5200 cm$^{-1}$ has completely disappeared in $\mathbf{A}_{DMS}$ while in $\mathbf{A}_{net}$ curves shown in figure 2, the absorption peak of water at this wavenumber has not been removed yet indicating the insufficiency of the subtraction method. A prominent peak at 5100 cm$^{-1}$, in the $\mathbf{A}_{DMS}$ has been observable. As illustrated in Figure 4 as the pure glucose absorbance, this peak can be due to glucose peak absorption at this wavenumber. The glucose solution transmittance was measured from a glucose-water solution by the FT-IR device. The concentration of the glucose in this solution was set as 100 [mg/dL]. Then, the pure glucose absorbance was calculated by subtracting the pure water absorbance from the calculated glucose solution absorbance.

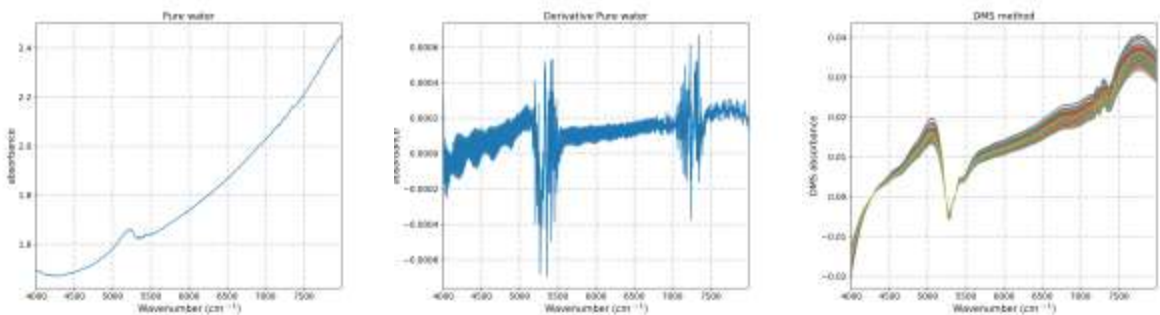

Figure 3. the pure water absorbance (left), its first derivative (middle) and the obtained absorbances after applying derivative multiplication suppression ($\mathbf{A}_{DMS}$) (right).

The first derivative of the glucose absorbance with respect to the wavenumber is also depicted in Figure 4. In Figure 5, (1-first derivative of the glucose absorbance) and the resulted DMR absorbances are illustrated. As can be observed, a distinguishable peak has been emerged a wavenumber close to 5100 cm$^{-1}$ that should be attributable to glucose. Therefore, DMR can work together with DMS to enhance the extraction of glucose peaks.

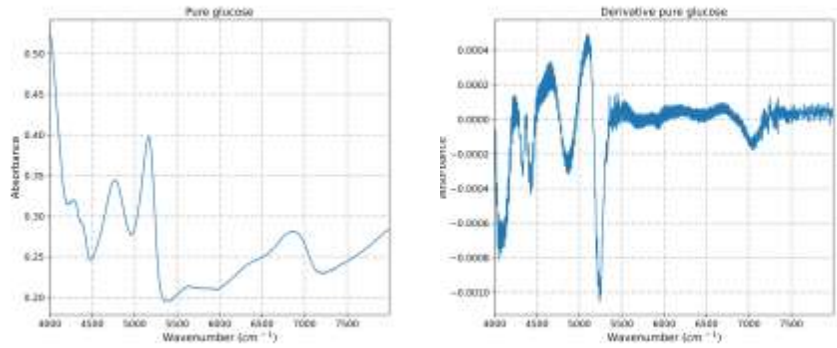

Figure 4. Pure glucose absorbance calculated from the glucose measured transmittance by the FT-IR device (left). The first derivative of the glucose absorbance (right).

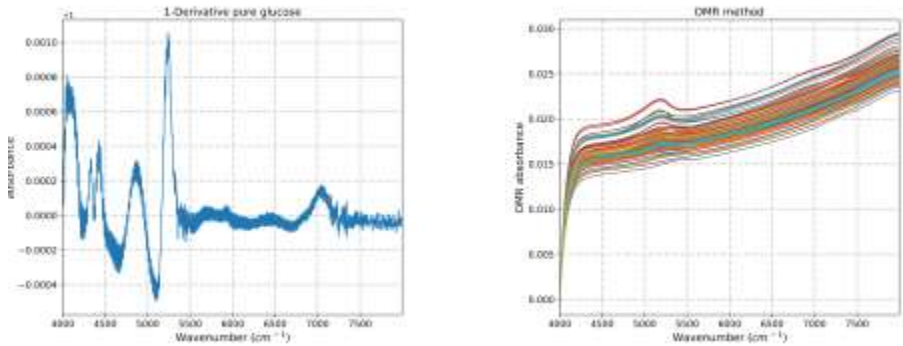

Figure 5. 1-first derivative of pure glucose absorbance (left) and DMR absorbance (right) spectra.

In Figure 6, the DMSR spectra are plotted. It is clear that the peaks at wavenumber 5100 are more distinguished in comparison to net absorbance, DMS, and DMR techniques.

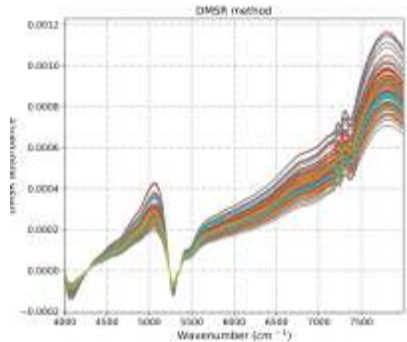

Figure 6. DMSR spectra

In Figure 7, the standardized forms of $\mathbf{A}_{net}$ ($\gamma$=1) and $\mathbf{A}_{DMS}$ have been shown and it is clear that the absorbance spectra have been less affected by the standardization in the DMS case compared to the net case. The standard form of the absorbances is employed for PCA. This invariability of the standardized form of DMS may be very advantageous to the multivariate analysis.

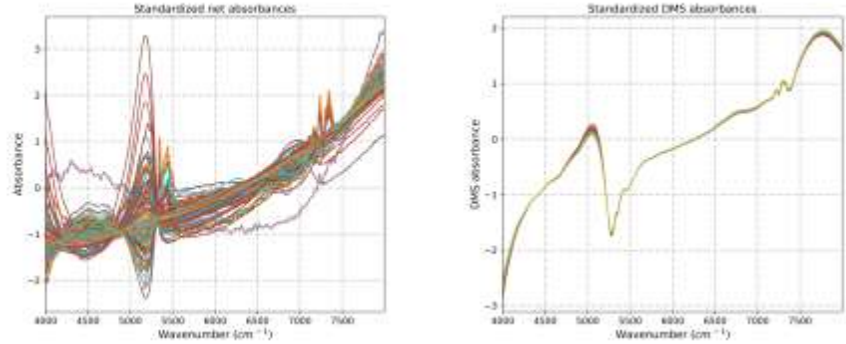

Figure 7. Net (left) and DMS (right) absorbances after standardization.

Then, the molar absorptivity of glucose has been calculated considering the Lambert-Beer relationship, viz. $\mathbf{A}(\tilde{\upsilon},\mathbf{c}) = \eta(\tilde{\upsilon})\mathbf{c}l$. $\mathbf{c}$ is the concentration of the solute (glucose). $l$ is the sample thickness. $\eta(\tilde{\upsilon})$ is the molar absorptivity of the solute which is a function of the wavenumber $\tilde{\upsilon}$. As described previously in the ARA approach, the molar absorptivity of glucose can be used for the direct calculation of the regression coefficients, which can then be used to estimate the sample glucose concentration from the measured absorbance. Our ARA model has provided glucose level predictions with an RMSE of about 22.42 [mg/dL], which can be considered an acceptable error compared to errors reported in other similar researches.[22,29] In figure 8, the measured and the ARA estimated blood glucose levels have been plotted for $\mathbf{R}_{n\times 1} = 105$. With this value, the least error or the best accuracy is achievable for the glucose level prediction of our measured data.

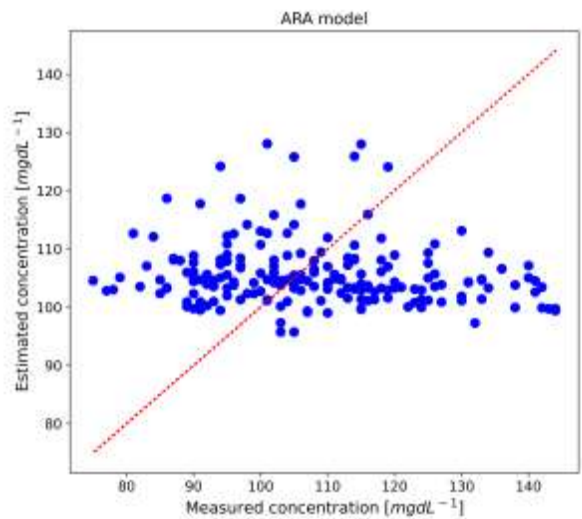

Figure 8. ARA predicted blood glucose concentrations vs. actual values. The line represents the ideal case at which the prediction perfectly follows the measurement: 4000-8000 cm$^{-1}$ (left), 4000-6000 cm$^{-1}$ (middle), and 4000-5500 cm$^{-1}$ (right)

For the PCR analysis, the empirical data has been divided into two groups, one as the training data and the other as the testing data. Our preliminary investigations showed that the best accuracy is obtained when 90% of the data is taken as the training set and the remaining 10% of the data is chosen for the test group. All the stages of PCR analysis have been implemented using the relevant functions provided by the Python package. The PCR analysis have been performed in three different ranges of wavenumber, i.e., I: 4000-8000, II: 4000-6000 and III: 4000-5000 cm$^{-1}$, in order to investigate the effect of the wavenumber windowing on the resulted data. In figure 9, the RMSE dependence on the number of PCs for the net, DMS, DMS, and DMSR cases at the data range I, are shown. As can be seen, when the net absorbances are selected for PCR, the least RMSE of about 13.14 [mg/dL] is generated when the number of PCs is 12. On the other hand, when the DMS absorbances are employed, the minimum RMSE occurs at PCs number equal to 10. In this case, RMSE is 12.5 [mg/dL]. For the case of DMR absorbances, the PCR error becomes the least equal to 12.7 [mg/dL] at PCs number of 3. Finally, PCR based on DMSR results in the smallest error of 12.77 [mg/dL] with PCs number of 8. The summary of the metrics of the models are listed in Table 1. The PCR-predicted blood glucose concentrations versus actual values (measured concentrations) are also shown for this wavenumber range. It can be seen that the DMS technique provides more accurate results in this range compared to the other methods. The fitted lines for the techniques of NET, DMR, and DMRS overlap. In Figures 10 and 11, the results of the same study at ranges of II and III are depicted.

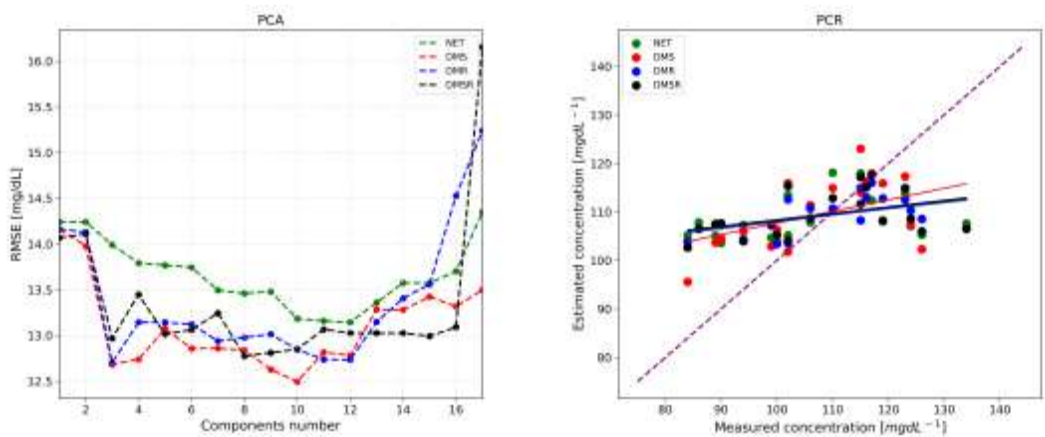

Figure 9. (Wavenumber range: 4000-8000 cm$^{-1}$) RMSE versus the number of the principal components for the cases that the net absorbances and the derivative-based absorbances have been employed for the PCR analysis (left). PCR predicted blood glucose concentrations vs. actual values. The dashed line represents the ideal case at which the prediction perfectly follows the measurement. The fitted lines corresponding to each case are also shown as solid lines (right).

Table 1. Metrics of the predictive approaches calculated in various wavenumber ranges.

| Model | RMSE [mg/dL] | MAE [mg/dL] | Standard Deviation of CV Scores [mg/dL] | PCs number |
|---|---|---|---|---|
| | I/II/III | I/II/III | I/II/III | I/II/III |
| NET | 13.14/13.62/13.62 | 10.77/10.92/11.19 | 28.03/104.27/46.37 | 12/11/11 |
| DMS | 12.50/12.95/12.80 | 9.73/10.65/10.38 | 17.17/3.51/3.02 | 10/7/5 |
| DMR | 12.70/12.77/12.97 | 10.09/10.12/10.29 | 3.48/3.85/4.12 | 3/3/3 |
| DMSR | 12.77/13.00/13.00 | 10.30/10.66/10.64 | 2.81/3.38/2.67 | 8/8/6 |
| ARA | 22.42/ - / - | 15.05/ - / - | - | - |

I: 4000-8000 cm$^{-1}$, II: 4000-6000 cm$^{-1}$, and III: 4000-5500 cm$^{-1}$.

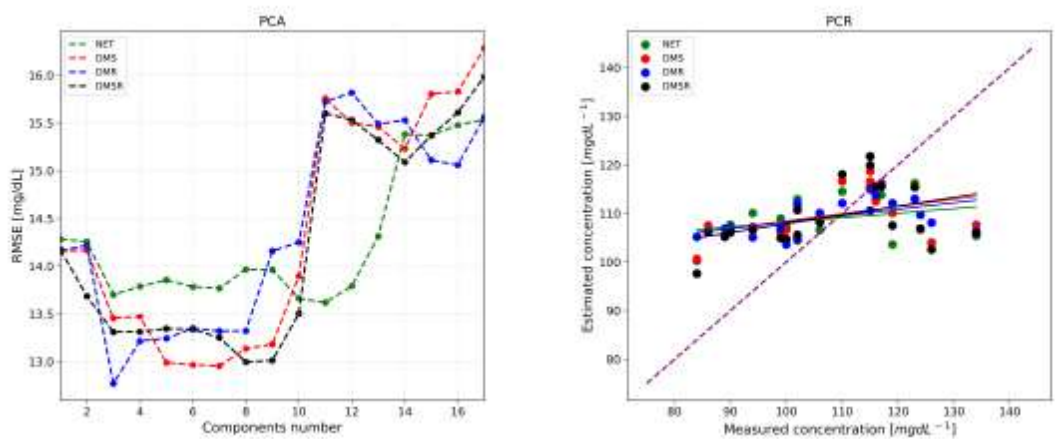

Figure 10. (Wavenumber range: 4000-6000 cm$^{-1}$) RMSE versus the number of the principal components for the cases that the net absorbances and the derivative-based absorbances have been employed for the PCR analysis (left). PCR predicted blood glucose concentrations vs. actual values. The dashed line represents the ideal case at which the prediction perfectly follows the measurement. The fitted lines corresponding to each case are also shown as solid lines (right).

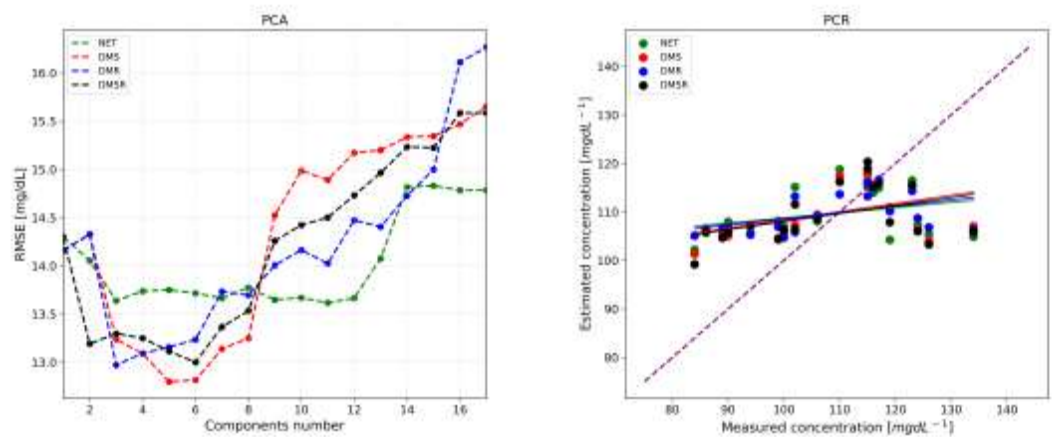

Figure 11. (Wavenumber range: 4000-5500 cm$^{-1}$): RMSE versus the number of the principal components for the cases that the net absorbances and the derivative-based absorbances have been employed for the PCR analysis (left). PCR predicted blood glucose concentrations vs. actual values. The dashed line represents the ideal case at which the prediction perfectly follows the measurement. The fitted lines corresponding to each case are also shown as solid lines (right).

From the results collected in Table 1, it can be deduced that when the net absorbances are utilized for PCR-based glucose predictions, the RMSE and Mean Absolute Error (MAE) values are slightly different in the three ranges. The number of PCs is roughly the same. However, the standard deviation of Cross-validation (CV) scores varies largely. It is the highest for range II and the lowest for range I. It can be inferred that if the net absorbances are employed for PCR-based glucose estimation, it is better to use the entire absorbance spectrum because a smaller value of the standard deviation of CV scores indicates that the predictive model exhibits more stability and robustness. In Table 1, it can be observed that the RMSE and MAE values are slightly reduced by using DMS, DMR, and DMSR compared to the NET case. However, the standard deviation of CV scores is much smaller than that of the NET case. In the DMS case, this value is the lowest when range III is used, suggesting that glucose information can be more effectively extracted when that portion of the wavenumber range near the expected glucose absorption peak is utilized. Nevertheless, although a similar value for the deviation is obtained for all three ranges in each case of DMS, DMR, and DMSR, DMSR provides the lowest value. Hence, it can be deduced that if the DMSR method is employed, glucose level estimation can be achieved in a more stable and robust manner with the least prediction error among the other techniques. However, DMR provides a simpler model due to its smaller number of PCs. For the ARA analysis, we have found that if only a portion of the spectrum is used instead of the entire spectrum, the accuracy obtained is poor. Therefore, neither the Standard Deviation of CV Scores nor the number of PCs is useful. Although the ARA model exhibits the lowest accuracy among other models, it can still be utilized for glucose prediction with acceptable accuracy because its error is not much larger than the other methods. However, The ARA technique stands out for its simplicity in implementation compared to other methods, as it relies solely on a linear regression approach with predefined linear coefficients. This makes it more accessible and easier to apply.

To evaluate the accuracy of blood glucose level predictions using the applied methods, the Clarke error grid analysis is typically employed. In Figure 12, the Clarke error grid analysis is illustrated for all approaches for the glucose level prediction. It can be observed that all methods demonstrate remarkable accuracy. Specifically, most of the PCR predicted values fall within region A, indicating prediction errors of less than 20%. Additionally, the predictions by the ARA model show the majority falling within region A, with a few falling within region B, indicating errors larger than 20% but still providing appropriate estimation.

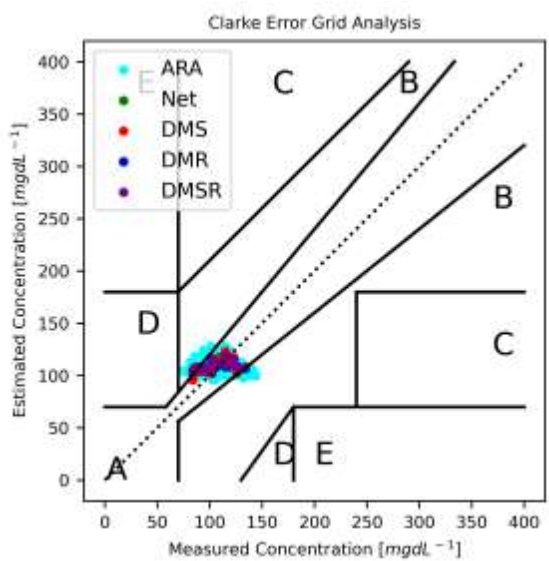

Figure 12. Clarke error grid analysis of the ARA model and the PCR approaches.

**Conclusion**

Blood NIR absorbances were recorded using the FT-IR approach. It was demonstrated that multiplying the absorbance curves by the first derivative of the water absorbance spectrum effectively removes peaks related to water interference in the blood absorbance spectra. Additional multiplication by the derivative of the glucose absorbance showed improved accuracy in the prediction of blood glucose. The PCR prediction using NET, DMS, DMR, and DMSR methods showed different accuracy in different wavenumber ranges. The derivative-based methods provided prediction results with smaller errors in comparison to the NET case with less sensitivity to the wavenumber-range variation. In addition, fewer principal components in the derivative methods were needed. Additionally, predicting blood glucose levels using the ARA regression model also showed acceptable accuracy, i.e., an error of 22.4 [mg/dL].

**Data accessibility.** This article has no additional data.
**Declaration of AI use.** We have not used AI-assisted technologies in creating this article.
**Authors 'contributions.** H.B.: writing—original draft, writing—review and editing; A.M.M.: writing—original draft, writing—review and editing; S.M.: writing—original draft, writing—review and editing; and editing M.F.: writing—original draft, writing—review and editing.
**Conflict of interest declaration.** We declare we have no competing interests.
**Funding.** This work has not been supported financially by anyone or any organization or any institute.
**Data availability statement.** The data that support the findings of this study are available within the article.